\documentclass{svmult} 

\usepackage{graphicx}        
\input{psfig.sty}
\usepackage{graphics}

\newcommand{\ar}{\arrowvert}
\newcommand{\ra}{\rangle}
\newcommand{\la}{\langle}
\newcommand{\da}{\dagger}

\newcommand{\cd}{\! \cdot \!}
\newcommand{\be}{\begin{equation}}
\newcommand{\ee}{\end{equation}}
\newcommand{\ba}{\begin{eqnarray}}
\newcommand{\ea}{\end{eqnarray}}

\begin{document}

\title*{RPA Vector Meson Leptonic Widths}
\author{Felipe J. Llanes-Estrada\inst{1}\and Stephen R. Cotanch\inst{2}}
\institute{Univ. Complutense de Madrid, Depto. F\'{\i}sica
Te\'orica I, 28040 Madrid, Spain
\texttt{fllanes@fis.ucm.es}
\and North Carolina State University, Dept. of Physics,
Raleigh, NC 27695, USA \texttt{cotanch@ncsu.edu}}

\maketitle

Recent $J = 0$ (para) baryonium interpretations of the BES narrow resonance
data
near the $e^-e^+ \rightarrow p \bar {p}$ threshold suggests the existance
of ortho  baryonium. To assist future  searches we study
$J = 1$ states, especially vector meson leptonic decays,
and report RPA calculations for both light and heavy mesons
using a   Coulomb-gauge
QCD-inspired model.  Since the $\phi(1880)$ is the only
missing  model state, other
discovered $J = 1$ particles in this region are  ortho  baryonium candidates.

\section{Introduction}
\label{sec:1}
Since the {\it {November Revolution}} entailing the discovery of charmonium
and universal acceptance of quarks, electron-positron collisions
have been an effective method for novel hadronic production.
The narrow widths of the $J/\psi$ and subsequently observed $\Upsilon$ led
to their
interpretation in terms of  new quark flavors (see for example
\cite{appelquist}).
These mesons were quarkonium states which predominantly decayed via the
electroweak interaction.
In a potential quark model their leptonic widths were first
calculated~\cite{schnitzer} using
\be \label{nonrelwidth}
\Gamma_{ee} \equiv \Gamma_{V \to e^-\ e^+} = 16 \pi \alpha_{\rm EM}^2
\frac{q^2}{M_V^2}
\ar \psi(0) \ar^2
\ee
with $q$ the charge of the quark in electron units, $M_V$ the resonance
mass and $\psi(0)$ the wavefunction at the origin.
The quantity $\ar \psi(0) \ar^2$ also appears in the matrix element of the
hyperfine interaction  \cite{appelquist} and therefore its model value
for the hyperfine splitting, $\Delta M_{hyp}$, in meson spectra
(pseudoscalar-vector meson mass differences)
should be simultaneously tested.
This also provides
a relation between the leptonic width and the
hyperfine splitting
\be
\Delta M_{hyp}= \frac{1}{6}\left(
\frac{9}{\alpha_{EM}^2} \frac{\Gamma_{ee }}{M_V}
\right)^{4/3}M_V \ .
\ee
However, using this result and the known $\pi$-$\rho$ mass difference, the
first prediction
\cite{appelquist} for the charmonium hyperfine splitting failed dramatically.
Subsequently, quark
models, such as Ref. \cite{isgur}, incorporating a complex, flavor
dependent hyperfine
interaction  were able to reproduce the hyperfine
splittings in various circumstances but
fine-tuning was still required since these approaches did not
include chiral symmetry.  The key point is that the pion is the
Goldstone
boson of  spontaneously broken chiral symmetry which is responsible for its
light mass.  Indeed more recent and improved chiral approaches
\cite{lc,weall} find that
about 70 \% of the $\pi$-$\rho$ splitting is due to
chiral symmetry.  Consequently, attributing this large 600 MeV
splitting entirely
to the hyperfine interaction can lead to an inconsistent description of
other, less dramatic
spectra splittings in baryons, heavy mesons and light meson excited states.

A more fundamental treatment is provided by field-theoretical quark models that
dynamically incorporate chiral symmetry. The Nambu and
Jona-Lasinio model  is one example but the signature contact
interaction precludes describing radial excitations (as well as
confinement). A similar
problem plagues contemporary covariant formulations of the Schwinger-Dyson
and Bethe-Salpeter equations that generate unphysical
states corresponding to relative time excitations. A more promising
approach, which we have utilized \cite{weall}, is to
diagonalize an effective QCD  Hamiltonian
in the Coulomb-gauge  using the chiral symmetry preserving Random
Phase Approximation [RPA].

In this paper we apply the RPA to predict vector meson
leptonic widths. In addition to further testing our model by confrontation with
data we also discuss exotic
baryonium states.  In view of the recent resonance discovered at
BES~\cite{bes} with a nucleon-antinucleon bound state interpretation
\cite{gao,kerbikov},
we comment on possible vector states in the $1800\ MeV$ region,
focusing on the quark model's, as yet undiscovered, $\phi(1880)$ that
naturally fits in our approach.

\section{Vector Meson Self-Energy and Width}

The width, $\Gamma$, of a hadron is inversely related to its lifetime and
can be obtained from its self-energy, $\Sigma (p^2)$, by
$Im \ \Sigma(M^2) = \Gamma/2$ for a particle with mass $M$ and momentum $p$.
For a vector meson the appropriate self-energy contribution
at the one $e^- e^+$ loop level can be computed  from
the diagram in Fig. 1.  This yields
\be
{\Sigma}=i \frac{g(M^2)^2}{M^2} \Pi(M^2)
\ee
where $g(p^2)$ is the meson-photon transition form factor and
$\Pi(p^2)$ is obtained from the photon polarization tensor with form
governed by gauge
invariance
\be
\Pi_{\mu\nu}(p^2)=(g_{\mu\nu}p^2 - p_\mu p_\nu) \Pi(p^2) \ .
\ee

\label{sec:2}
\begin{center}
\begin{figure}[h]
\hspace{2cm}
\psfig{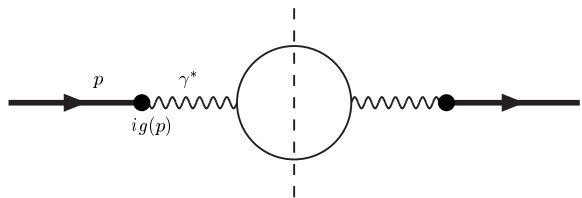}
\caption{\label{fig1} Feynman diagram for the one loop $e^- e^+$
contribution to the self-energy of a meson (solid line) with momeum $p$.
All hadron information is contained in the transition form factor $g(p)$.}
\end{figure}
\end{center}

Applying standard QED covariant perturbation
theory to evaluate the photon
polarization tensor yields
\be \label{tensor}
\Pi_{\mu\nu} \simeq -e^2 \int \frac{d^4 q}{(2\pi)^4} \frac{{\rm Tr}(
\gamma_\mu \not q \gamma_\nu (\not q +\not
p))}{(q^2+i\epsilon)((q+p)^2 + i\epsilon)}  \ .
\ee
Note we have neglected the electron's mass since it is a small effect.
Related, Eq. (\ref{tensor}) also represents the one loop $\mu^- \mu^+$
contribution
in the zero mass limit and, as suggested by Table 1, this mass can also be
neglected
since $\Gamma_{\mu\mu}$ is essentially equal to $\Gamma_{ee}$ within
experimental error.
This conclusion is
reinforced by $\Gamma_{\mu \mu}$ being randomly larger or smaller
than $\Gamma_{e e}$, since phase space would indicate it to be
smaller. Further, the decay width for $\Upsilon\to \tau^-\tau^+$ is also known
and is
compatible with the widths for $\mu^- \mu^+$ and $e^-e^+$ as
well.
The effect of the $\mu$ mass should be maximal for the $\rho$ and $\omega$
widths
where the momentum of the outgoing particles is not as large. The relevant
correction (see discussion below), $4m_\mu^4/p^4$,  produces a $10\ \%$
suppression
which is less than experimental error.
\begin{table}
\begin{center}
\caption{Observed vector meson masses and leptonic widths.
Partial widths estimated from the central PDG~\cite{pdg} value for the
total width.  The error quoted, affecting the last digit, corresponds
to the uncertainty in the branching ratio alone.
 \label{emu}}
\begin{tabular}{cccc}
State &$M(MeV)$& $\Gamma_{ee}(keV)$ & $\Gamma_{\mu \mu}(keV)$\\
\hline
$\rho$         & 770   &  6.81(2)  &  6.9(4)   \\
$\omega$       & 783   &  0.60(2)  &  0.76(25) \\
$\phi$         & 1019  &  1.26(2)  &  1.22(9)  \\
$J/\psi$       & 3097  &  5.2(1)   &  5.1(9)   \\
$\psi(2S)$     & 3686  &  2.1(1)   &  2.1(3)   \\
$\Upsilon$     & 9460  &  1.32(5)  &  1.30(3)  \\
$\Upsilon(2S)$ &10023  &  0.52(3)  &  0.58(9)  \\
\hline
\end{tabular}
\end{center}
\end{table}

Performing the Dirac traces in Eq. (\ref{tensor}) and employing the tensor
integral relations for a scalar function $F$
\ba
\int  q_\mu q_\nu F(q,p) \  d^4 q &=& \frac{g_{\mu\nu}}{3}  \int [q^2 -
\frac{(q\cd p)^2}{p^2}]  F \  d^4 q  \nonumber \\
&+& \frac{p_\mu p_\nu}{3p^4}\int [4(q\cd p)^2 -p^2  q^2] F \ d^4 q \\
\int q_\mu F \  d^4 q &=&\frac{p_\mu}{p^2} \int q\cd p F \  d^4 q
\ea
one can obtain scalar integral expressions. Since we
only require the imaginary (finite) part of the $ \Sigma$
self-energy, we can ignore all real (known to be UV divergent)
contributions. For
example, we suppress the tadpole integral
$$
\int \frac{d^4 q}{(2\pi)^4} \frac{q^2}{(q^2+i\epsilon)((q+p)^2+i\epsilon)}
=\int \frac{d^4 q}{(2\pi)^4}\frac{1}{(q+p)^2+i\epsilon}\to 0
$$
in this calculation (the real, mass shifts from lepton loops are totally
negligible).
After some manipulation
\be
{\rm Im} \  (i \ \Pi(p^2)) = - \frac{4e^2}{3} \ {\rm Im} \ ( i \ J)
\ee
where $J$ is the standard massless boson loop integral
\be \label{bubble}
J= \int \frac{d^4 q}{(2\pi)^4}
\frac{1}{(q^2+i\epsilon)((q+p)^2 +i\epsilon)} \ .
\ee
The $q_0$ integral is easily performed in the complex plane and the
remaining integral can be obtained with the
substitution
$$
\frac{1}{E-E_r-i\epsilon}\to 2\pi i \delta(E-E_r) \ .
$$
The result (identical to using Fermi's golden
rule) is
$$
{\rm Im} (i \  J) = -\frac{1}{16\pi}
$$
giving the theoretical leptonic width
\be \label{imself}
\Gamma = 2 \  {\rm Im} {\Sigma}(M^2) =  \frac{2\alpha_{EM}}{3M^2} g^2(M^2) \ .
\ee
For completeness, the correction to Eq. (\ref{imself}) due to a non-zero
lepton mass, $m$, in the polarization tensor  is obtained \cite{peskin}
by replacing the explicit
fine structure constant with
\ba
\alpha_{EM} \to \alpha_{EM} \sqrt{1-\frac{4m^2}{q^2}}\left(
1+\frac{2m^2}{q^2} \right)
\simeq \alpha_{EM} \left( 1-\frac{4m^4}{q^4} \right) \ .
\ea

\section{RPA Leptonic Widths}

To evaluate $\Gamma_{ee}$ in the RPA we need to calculate the
transition form factor $g(p)$.  As indicated in Fig. 2  there are both
meson and quark
loop contributions.
\begin{figure}[h]
\psfig{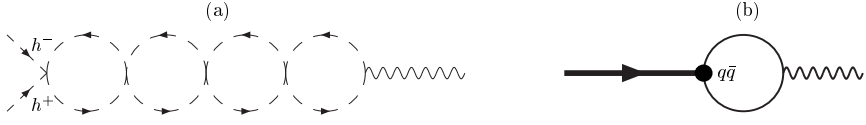}
\caption{
\label{fig2} Diagrams for the
transition form factor $g(p)$: (a) hadron decay and rescattering before
annihilation;
(b) direct quark-antiquark annihilation.}
\end{figure}
In the next section we show that the quark-antiquark annihilation, Fig. 2b,
is the
dominant process with contribution in the RPA given by
\be \label{rpawidth}
g = \frac{eq}{\la \Omega \ar \Omega \ra}
\la \Omega \ar \left[ \Psi^\da (0) \alpha^i \Psi(0), V^{i  \da }
\right] \ar \Omega \ra_{\rm RPA}
\ee
where $\Psi$ is the quark field operator and $q$  the  (fractional) quark
charge
in electron units.  The RPA vector
meson creation operator with spin projection $i$ is denoted by $V^{i  \da}$ and
contains creation and annihilation parts
\be
V^{i  \da}= X^i_{\alpha\beta} B^\da_\alpha D^\da_\beta-
Y^i_{\alpha \beta} B_\alpha D_\beta
\ee
with all remaining quantum numbers represented by $\alpha$ and $\beta$.
For  $J^{PC}=1^{--}$  states there are $S$ and $D$ wavefunction
components
\ba
X_S^i({\bf k})&=&\frac{\delta^{ab}}{\sqrt{N_c}} \frac{1}{\sqrt{4\pi}}
\frac{\sigma^i i \sigma_2}{\sqrt{2}} X_S(k) \\
Y_S^i({\bf k})&=&\frac{\delta^{ab}}{\sqrt{N_c}} \frac{-1}{\sqrt{4\pi}}
\frac{i \sigma_2 \sigma^i}{\sqrt{2}} Y_S(k) \\
X_D^i({\bf k})&=&\frac{\delta^{ab}}{\sqrt{N_c}} \frac{1}{\sqrt{4\pi}}
\frac{3}{2} \left(\hat{\bf k}\cd{\bf \sigma}\hat{k}^i-\frac{1}{3}\sigma^i
\right)i\sigma_2 X_D(k) \\
Y_D^i({\bf k})&=&\frac{\delta^{ab}}{\sqrt{N_c}} \frac{-1}{\sqrt{4\pi}}
i\sigma_2\frac{3}{2} \left(\hat{\bf k}\cd{\bf
\sigma}\hat{k}^i-\frac{1}{3}\sigma^i
\right)Y_D(k)
\ea
with color indices $a,b,c$ and $N_c = $ 3.
The normalization condition
\be
\frac{\la \Omega \ar \left[ V^i,V^{i\da}\right] \ar \Omega\ra }{\la
\Omega \ar \Omega \ra} = 1
\ee
reduces to
\ba \label{norm}
\frac{1}{(2\pi)^3} \int_0^\infty k^2 dk
\left( \ar X_S(k)\ar^2
+\ar X_D(k)\ar^2-\ar Y_S(k)\ar^2-\ar Y_D(k)\ar^2 \right) = 1
 \ .
\ea

The scalar wavefunctions are solutions to coupled integral equations
resulting from the RPA diagonalization of the Coulomb gauge
Hamiltonian~\cite{weall}.  A state-of-the-art QCD confining potential
with  scale (string tension or gluon generated mass) connected to
$\Lambda_{\rm QCD}$ was used along with
a transverse gluon exchange hyperfine interaction, essentially a Yukawa
potential, supported by recent lattice results \cite{langfeld}, with a
range parameter chosen to adjust  the mass-spectra.

In evaluating Eq. (\ref{rpawidth}) we note that spin is conserved in the
vector meson-photon transition
(diagonal coupling). Rotational invariance permits summing over the
three spin components if we divide by 3.
The resulting RPA $g$ is then
\ba \label{qqtophoton}
g &=&\frac{eq\sqrt{8\pi N_c}}{3} \int_0^\infty \frac{k^2dk}{(2\pi)^3}
\Big [  (X_S(k)-Y_S(k)) (2+\sin \phi(k))   \nonumber
\\
&& -\sqrt{2}(1-\sin\phi(k))
(X_D(k)-Y_D(k)) \Big ] \ .
\ea

Note that the RPA leads to an expression of the form given by Eq.
(\ref{nonrelwidth}) since $\Gamma\propto g^2$, and
in the non-relativistic limit
\be
g \to e q \sqrt{8\pi N_c} \int_0^\infty  \frac{k^2dk}{(2\pi)^3}
X_S(k)
\ee
because
 $sin \  \phi(k)$, the BCS vacuum angle (solution of the gap
equation for the same Hamiltonian), approaches unity
\be
\sin \phi(k)= \frac{M_q(k)}{\sqrt{k^2+M_q(k)^2}} \simeq 1
\ee
for large running quark mass and the $Y$ function can be neglected
in the same limit.
Finally the Fourier-transformed wavefunction is
\be
\psi(0)\propto \int e^{ik\cd 0} \frac{d^3k}{(2\pi)^3}
\frac{X_S(k)}{\sqrt{4\pi}} \ .
\ee

The effective charge factors, $q^2 \rightarrow <q^2>$, for the various
quark flavors necessary for
our width calculation are listed in Table \ref{flavor}.
This is the origin of the order of magnitude suppression of the $\omega$
width relative to the
$\rho$. We work in the
exact isospin limit.
\begin{table}[h]
\begin {center}
\caption{Quark charge factors $\la q^2\ra$.
\label{flavor}}
\begin{tabular}{c|c|cccc} State  & Flavor  & & &
\multicolumn{2}{c}{$\la q^2\ra$}
 \\
\hline
$\Upsilon$ & $b\bar{b}$ & & & $\frac{1}{9} $& \\
$J/\psi$ & $c\bar{c}$ &  & & & $\frac{4}{9}$ \\
 $\phi$ & $s\bar{s}$ & & & $\frac{1}{9}$& \\
$\rho$ & $\frac{u\bar{u}+d\bar{d}}{\sqrt{2}}$ & & & & $\frac{1}{2}$  \\
$\omega$ & $\frac{u\bar{u}-d\bar{d}}{\sqrt{2}}$ & & & $\frac{1}{18}$ & \\
\hline
\end{tabular}
\end {center}
\end{table}

Combining Eqs. (\ref{imself}) and (\ref{qqtophoton})
yields the final expression for the  RPA  widths.  The
numerical predictions for both meson masses and widths are listed in Table
\ref{widths}.
The predicted
masses are only in rough agreement since a fit was not attempted (the quark
masses were not
fine-tuned).
To avoid compounding errors, the physical
meson masses were  used in Eq. (\ref{imself})
except for the unknown $\phi(1880)$ where the model mass is adopted.
Given the large errors in the measured lepton widths this should be a minor
concern.  For
comparison, two other theoretical calculations~\cite{isgur,ebert} are also
presented.
It is noteworthy that our few parameter approach provides
a description which is comparable to these multi-parameter models.

\begin{table}
\begin {center}
\caption{Experimental and theoretical vector
meson widths.  Also listed are two other calculations, marked
$^1$ (Godfrey and Isgur \cite{isgur} $S$-$D$ wave
results without mixing) and  $^2$
(Ebert, Faustov and Galkin  \cite{ebert}).
 \label{widths}}
\begin{tabular}{c|cc|ccc}
State & Exp. $M(MeV)$ &Calculated & Exp. $\Gamma_{e^- e^+ }(keV)$ &
Calculated & Other works \\ \hline
$\rho$     &  770      & 795 & 6.85     & 4.7 & $6.87^1$ \\
$\omega$   &  782      & 795 & 0.60(2)  & 0.50 & $0.85^1$ \\
$\phi$     &  1019     & 1005 & 1.26(2)  & 1.1 & $2.75^1$ \\
$\omega$   &  1420     & 1420& $0.1(1)^\da$&0.18&
$0.046^1$\\
$\rho$     &1450(25)   & 1420& -        & 1.5 & $0.45^1$\\
$\omega$   &1650(25)   & 1620& -        & 0.23 & $0.013^1$\\
$\phi$     &1680(25)   & 1670& $0.5(10)^\da$& 0.21 &
$0.27^1$\\
$\rho$     &1700       & 1520&$3.5(5)^\da$& 2.0 &
$0.14^1$\\
$\phi$     & -         & 1790&   -      & 0.45-0.55 & $0.06^1$ \\
$J/\psi$   &  3097     & 3130& 5.2(1)   & 3.8 & $9.9^1$, $5.4^2$ \\
$\psi$     &  3686     & 3681 & 2.2(2)   & 2.0 & $3.3^1$,$2.4^2$ \\
$\psi$     &  3770     & 3695 & 0.26(4)  & 0.68 & $0.10^1$\\
$\psi$     &  4040     & 4140 & 0.75(15) & 1.7 & $1.8^1$\\
$\psi$     &  4160(20) & 4150 & 0.77(23) & 0.56 & \\
$\psi$     &  4415(6)  & 4535 & 0.47(10) & 0.26 & \\
$\Upsilon$ &  9460     & 9460 & 1.32(5)  &  0.45 & $1.4^1$,$1.3^2$ \\
$\Upsilon$ &  10023    & 9870 & 0.52(3)  &  0.35 & $0.65^1$, $0.5^2$ \\
$\Upsilon$ &  10365    & 9921 & $0.5(1)^\ddag$  &  0.0003 &$0.45^1$
\\
$\Upsilon$ &  10580    & 10190 & $0.32(3)^\S$ & 0.29 & $0.34^1$\\
\hline
\end{tabular}
\end {center}
{$^\da$our estimate} \ \ $^\ddag$
$\mu^+\mu^-$ width (roughly equal)\ \ $^\S$ recently reported by
BaBar \cite{babar}
\end{table}

Finally, we can relate our form factor parameter, $g$, to the standard
vector meson leptonic decay constant, $f_V$, defined by
\be \label{fvtogamma}
\Gamma_{e^- e^+}= \frac{4\pi}{3}\alpha^2_{EM} M_V f_V^2 \ .
\ee
Comparing to  Eq. (\ref{imself}) yields
\be
g=e f_V \sqrt{\frac{M_V^3}{2}}
\ee
with mass dimension consistent with Eqs.
(\ref{norm}) and (\ref{qqtophoton}). In representing the widths from
Ref.~\cite{isgur}  we use
their $f_V$ values and again available physical meson masses.
Combining the different models it is clear that we have a reasonable
theoretical understanding of
vector meson dileptonic decay.

\section{Mesons with Broad Hadronic Widths}

For vector meson decay to a particle-antiparticle hadron pair, annihilation
to a virtual photon is possible which also contributes to the total lepton
width.
In this section we will demonstrate that such  rescattering
corrections are small in comparison with the intrinsic quark-antiquark
annihilation contribution.
A typical example of a broad vector meson width is
the $\rho \rightarrow  \pi\pi$. We assume that the true eigenstate of
the full QCD Hamiltonian  also contains a two-pion wavefunction component
\be
\ar \rho \ra = \alpha \ar q\bar{q} \ra + \beta \ar \pi \pi \ra + \cd\cd\cd
\ee
with $\ar \beta \ar < 1$.
We have previously calculated  the decay of the $q\bar{q}$ component
into $e^-e^+$ and now proceed to estimate the contribution from the $\ar
\pi \pi \ra$ term.

In principle this higher Fock-space wavefunction component could be
computed from
the same model Hamiltonian
employed in the $q\bar{q}$ two-body sector since it is field-theory based.
However this is a challenging calculation involving four relativistic
particles and chiral symmetry. Although a formalism has
been developed \cite{pcfmrs}
to treat this problem, for an order of magnitude
estimate we instead implement a simple model wavefunction and
adopt the following ansatz for the $V = \rho$ two-pion
component
\ba \label{modelwf}
V^{i\ a  \da}(p=(M_V,{\bf 0})) \ar \Omega \ra &=&
\int
\frac{d^3k}{(2\pi)^3} \frac{\sqrt{3} \hat{k}^i}{\sqrt{4\pi}}
\frac{\epsilon^{abc}}{\sqrt{2}}
\frac{\Gamma_{\pi \pi}/2}{M_V-2E_k-i\Gamma_{V}/2}
\nonumber \\ &&
\sqrt{\frac{2(2\pi)^3}{2M_V kE_k}}\frac{1}{\sqrt{2}} \pi^{b \da}_{\bf k}
\pi^{c \da}_{-{\bf k}} \ar \Omega \ra \ .
\ea
Here $E_k=\sqrt{M_\pi^2+k^2}$, $\Gamma_V$ is the total $V$
width and $\Gamma_{\pi\pi}$ is the width for the $\pi^+\pi^-$ decay.
The indices $a, b$ and $c$ now denote charge states.
This wavefunction is inspired by the scattering
amplitude for $h^+h^-$ hadrons coupled to angular momentum 1
near a resonance described by the Breit-Wigner
amplitude
\be
a_{1}(s)=-\frac{M_V \Gamma_{h h}/2}{s-M_V^2+iM_V\Gamma_{V}}
\ee
or in the $V$ rest frame with $E = M_V + i \Gamma_V/2$
\be
a_{1}(E)=\frac{\Gamma_{h h}/2}{M_V-E-i\Gamma_{V}/2}
= \frac{i\Gamma_{h h}}{2\Gamma_{V}}  \ .
\ee
The normalization is given by
\ba
\frac{\la \Omega \ar\ \left[ V^{i\ a}, V^{j\ a' \da} \right] \ar \Omega
\ra}{\la \Omega \ar \Omega \ra} =
 \delta^{aa'} \delta^{ij} (2\pi)^3\delta^3(0) \ {\cal N}^2
\ea
where the  factor ${\cal N}^2$ (smaller than unity) is
\be
{\cal N}^2=\frac{1}{2M_V}\int_{2m_\pi}^\infty dE
\frac{\Gamma^2_{\pi\pi}/4}{(E-M_V)^2+\Gamma_{V}^2/4}
\ee
and represents the probability of finding a pion pair with
relative momentum given by the Breit-Wigner distribution. This determines
the $\beta$ coefficient above.
This normalization also represents the number of pion pairs per
$\rho$ under the Breit-Wigner curve.

The pion pair annihilation to a photon involves the electromagnetic
current
\be \label{picurrent}
j^{i\ a}(0)= i \epsilon^{abc} \pi^{b \da}(0) \partial^i \pi^{c \da}(0)
\ee
taken bare as all pion rescattering effects  are
already included in the Breit-Wigner distribution.
The pion field at the origin can be expressed in terms of the pion
momentum operator
\be
\pi^b(0)= \int\frac{d^3q}{(2\pi)^3}e^{i\vec{q}\cd\vec{0}} \pi^b_{\bf q}
\frac{1}{\sqrt{2E_q}} \ .
\ee
The hadronic transition coefficient is then
\be
g= eq \frac{1}{3} \sum_a \frac{1}{3} \sum_i \frac{\la \Omega \ar j^{i\
a}(0) V^{i\ a \da}\ar \Omega \ra}{\la \Omega \ar \Omega \ra}
\ee
where the charge factor is now 1 as both hadrons carry one electron unit
and spin and isospin are averaged over.
The  straightforward computation yields
\be
g=eq \frac{\sqrt{4\pi}}{\sqrt{3(2\pi)^3M_V}} \int_0^\infty
\frac{k^{5/2}dk}{E_k^{3/4}}
\frac{\Gamma_{\pi\pi}/2}{M_V-2E_k-i\Gamma_{V}/2} \ .
\ee
The imaginary part of the integral is finite, but the real part
diverges linearly with increasing momentum cut-off $\Lambda$. This reflects
short range physics and the need for a
counterterm in the vector current of Eq. (\ref{picurrent}).
Since  the ``contact''
part of the decay has been calculated  using the RPA, the counterterm
coefficient (or simply the
cutoff in the integral) could be fixed by the difference between the
physical  and
RPA lepton widths. But this generates an unnaturally large cutoff which is
not expected for pions with energies comparable to the mass $V$, i.e.
pion pairs with energies beyond a GeV from the resonance should not appreciably
contribute.  A third way to regularize the integral is to allow
for an energy-dependent width that localizes the integrand. In principle all
three ways are equivalent and we therefore use the simple
cut-off regularization with parameter $\Lambda$. The results for the
$e^-e^+$ widths corresponding
to the three lowest $\rho$  states  are presented in Table \ref{broad} for
different model parameters.
The momentum cut-off is listed dimensionlessly, $(\Lambda - M_V)/\Gamma_V$,
in terms of the $V$ mass
and width.

\begin{table}\caption{ Contribution to $\Gamma_{V \to e^-e^+}$ from $\pi \pi$
pion rescattering.
\label{broad}}
\begin {center}
\begin{tabular}{ccccc|c}
$M_V\ (MeV)$& $\Gamma_{V}\ (MeV)$ & $\Gamma_{\pi\pi}\ (MeV)$ &
$\cal N$& $(\Lambda-M_V)/\Gamma_{V}$ & $\Gamma_{ee}\ (keV)$ \\
\hline
700     & 150                       & 150                      & 0.38    &
5       & 0.068 \\
700     & 150                       & 150                      & 0.38    &
10      & 0.156 \\
1465    & 350                       & 100                      & 0.12    &
5       & 0.028 \\
1465    & 350                       & 100                      & 0.12    &
10      & 0.064 \\
1700    & 240                       & 80                       & 0.11    &
5       & 0.0070 \\
1700    & 240                       & 80                       & 0.11    &
10      & 0.014 \\ \hline
\end{tabular}
\end {center}
\end{table}

Note that the rescattering contribution
to the electron-positron widths are
suppressed relative to direct quark-antiquark annihilation  by an order
of magnitude. With this estimate and the present experimental width
precision, the RPA calculation appears to provide a sufficient
description without the need for rigorously including hadron final-state
interactions.
Further, rescattering effects for other meson decays can be even smaller.
For example,
in odd $G$ parity $\omega$-type meson decays to three or more (odd)
particle states,
the final state mesons are very unlikely to recombine and annihiliate to a
photon. Also the
$\phi$  and several other  quarkonium states have narrow widths
because they are near  or below hadronic decay thresholds. Finally, the
$\phi(1680)$ has an
$\omega$-like decay pattern with dominant $K^*K$ decay products  which
cannot directly
annihilate into a single photon.

\section{Towards baryonium}

The short-ranged (weak) $NN$ interaction, $V_{NN}$, can
only produce a single bound state, the deuteron.
In contrast, for the $N \bar{N}$ system the stronger $V_{N \bar{N}}$
interaction obtained from
$V_{NN}$ using $G$ parity can support several bound states.   However,
because of
annihilation and also mixing and decay to other baryon number zero hadrons
(mesons), it has long
been believed that baryonium states, if found, would be quite broad.
Significantly, BES has
reported~\cite{bes} a dramatic narrow enhancement in the decay $J/\psi\to
\gamma p\bar{p}$ but not
in
$J/\psi \to \pi_0 p \bar{p}$. This indicates the possibility of an
$I = 0$, as opposed to $I=1$, bound state not far below threshold
\cite{kerbikov}.
The quantum numbers of this state are likely to be either $J^{PC}=0^{-+}$
or $0^{++}$.  If it is a pseudoscalar state, it cannot be explained in a
$q\bar{q}$ model since
the expected excited $\pi$ and $\eta$ states in this region have already
been identified as the
$\pi(1800)$ and $\eta(1760)$.

The baryonium interpretation of this resonance suggests the pseudoscalar
assignment corresponding to \emph{para baryonium} (nucleon and antinucleon
spins antiparallel) which, in turn, implies a partner, \emph{ortho baryonium},
with spins alligned having total $J^{PC}=1^{--}$. A similar state is also
suggested by
the observed low-energy proton timelike form factor behavior as well as by
a $J^{PC}= 1^{--}$ resonance prediction using Vector Meson Dominance
\cite{williams}
just below
the $e^-e^+ \rightarrow p \bar{p}$ threshold.
Further, a state, with mass
$1870(10)\ MeV$ and width $10(5)\ MeV$, has also been reported
by the Fenice collaboration \cite{fenice} in $e^-e^+$ production  but
awaits confirmation. Another discrepancy with expected
phase-space is reported~\cite{abe} in $B\to \bar{p}pK$.

Since baryonium has conventional meson quantum numbers, it is difficult to
disentangle from $q \bar{q}$ and  $2q$-$2\bar{q}$
states. In particular, quark models predict numerous four-quark
states that likely overlap forming a  hadron continuum. The best prospect for
observing
baryonium states would be if they had narrow widths (possibly due to
significant quark
recombination) and  non-strange decay
products near the $N \bar{N}$ threshold. Such states are
predicted by the Resonating Group Method \cite{ribeiro}.
Because the  BES  and  FENICE
resonances both have widths of order $10\ MeV$, Coulombic baryonium states
bound by just QED
are unlikely to be observed  since these
would have very narrow leptonic widths given by
\be
\Gamma_{B\to e^-e^+} = \frac{4}{3}\alpha_{EM}^5 M_p/(2n)
\ee
that are of order $10\ eV$. Instead, we would anticipate one ortho and
one para
baryonium level each with a typical width of order 10 MeV.

Experimental information above threshold
should be regarded more cautiously.
As recently discussed~\cite{bugg},
steep peaks in the nucleon timelike form factors and $J/\psi$ decays could be
simple threshold cusp effects.
Consequently, baryonium claims should focus below threshold on
isolated peaks or dips in an observable like
the unconfirmed dip at 1870 MeV in the
total $e^-e^+$ annihilation cross section~\cite{fenice}.

\section{$\phi(1880)$}

Since ortho baryonium is a vector state it should be readily accessible
with new
high energy and luminosity $e^-e^+$ colliders,
such as the envisioned DA$\Phi$NE upgrade to 2 $GeV$ \cite{dafne}.
In this energy region, the most likely  hadron to be observed is
the excited $\phi$ meson, a combination of  $2S$ and $1D$ waves, with mass
$1880\ MeV$ predicted by
Isgur
and Godfrey  and $1790 \ MeV$ in our RPA model. As listed in Table
\ref{widths}, our calculated
$e^-e^+$ width for this state is about half a $keV$.

We can also phenomenologically estimate this width using
known data and flavor symmetry. Recall
Weinberg's first sum rule
\be \label{weinbergs}
\frac{1}{3} M_{\rho} \Gamma_{\rho\to e^- e^+} =
M_{\omega} \Gamma_{\omega\to e^- e^+} + M_{\phi} \Gamma_{\phi\to e^-
e^+}
\ee
which appears reasonable
\cite{desy} since
the left and right sides of this equation are 1.758 and 1.754 $MeV^2$,
respectively. Assuming ideal mixing for the $\omega$-$\phi$ system,
we obtain
\be \label{idealmix}
2M_\rho \Gamma_{\rho\to e^- e^+} = 9 M_{\phi} \Gamma_{\phi\to e^-
e^+} = 18 M_{\omega} \Gamma_{\omega\to e^- e^+}
\ee
which is satisfied to within $30\ \% $  ($10.5,\  11.6,\ 8.4 \
MeV^2$, respectively).  Combining Eqs. (\ref{weinbergs}) and (\ref{idealmix})
with the known
\cite{pdg} relative branching ratio
$$
\frac{\Gamma_{\rho(1700)\to e^-e^+} \Gamma_{\rho(1700)\to 2(\pi \pi)
}}{\Gamma_{\rho(1700)\to {\rm all}}}
$$
we obtain the estimates,
$\Gamma_{\rho(1700)\to e^-e^+}=3.5(0.5)\ keV$,
$\Gamma_{\phi(1880)\to e^-e^+}=0.7(3)\ keV$ and
$\Gamma_{\omega(1650)\to e^-e^+}=0.4(2)\ keV$.
These are somewhat larger but still in agreement with our model calculations.
Therefore, in searching for $J = 1$ baryonium candidates
below the $N{\bar {N}}$ threshold, there may be
a $\phi(1880)$ state with $\Gamma_{ee}$ about half a $keV$ which should be
distinguishable by
its
kaon decay modes.

\section{Outlook}
Recent claims have again raised the question concerning the existence of
baryonium
states but more solid information is needed near the $N{\bar {N}}$ threshold.
The upgrade of DA$\Phi$NE to the $2\ GeV$ range will provide a good
opportunity for further searches.  A possible fruitful
strategy would be to focus on $e^-e^+$ annihilation into
an even number of pions in isoscalar $1^{--}$ states.
Resonances, although not narrow, are anticipated and
there are structure precursors
\cite{baldini} from other experiments. By extracting the leptonic widths,
the missing $\phi$ state below 2 $GeV$ should be found
and we predict its lepton width to be around half a $keV$. Its hadron width
should be similar to the $\omega(1650)$, about 200 $MeV$.
Any other observed vector state is a candidate for ortho baryonium.

\emph{This work is supported by grants FPA 2000-0956, BFM 2002-01003 (MCYT,
Spain)
and DE-FG02-97ER41048 (US Department of Energy) and has
been commissioned for the ``Encuentro de
F\'{\i}sica Fundamental Alberto  Galindo'' meeting at Univ. Complutense,
November,
2004. The authors  are grateful to the workshop organizers and congratulate
Professor Galindo on
his Iubilaeum.}

\end{document}